\documentclass{kluwer}

\usepackage{epsfig}

\begin{document}



\begin{article}

\begin{opening}

\title{Multi-component models for disk galaxies.}
\subtitle{A test case on NGC 5866}
\author{E. \surname{Pignatelli}\email{pignatelli@pd.astro.it}}
\institute{Osservatorio Astronomico di Padova, vicolo dell'Osservatorio 5, I 35122 Padova, Italy}
\begin{abstract}

We present an application of a new set of detailed, self-consistent,
dynamical models for disc galaxies. We start from the hypothesis that
each galaxy can be decomposed in a bulge, following the $r^{1/4}$ law,
and a disc with an exponential projected density profile; and that the
isodensity surfaces of each component can be represented by similar
concentric spheroids. After taking into account both the asymmetric
drift effects and the integration along the line of sight, we produce
the rotational velocity and velocity dispersion profile, {\bf and }
the approximate shape of the line of sight velocity distributions for
the stars as parameterized by the $h_3$ and $h_4$ coefficients of the
Gauss-Hermite expansion of the line profile.

Photometric and kinematical data have been taken from the literature
for the test case of the S0 galaxy NGC 5866, for which detailed
stellar kinematical data are available at different positions across
the galaxy.  Apart from the very inner, dust-obscured regions of the
galaxy, where observational effects are likely to be dominant, the
model successfully reproduce the whole set of dynamical data available
as well as giving a good fit to the photometry. The galaxy is shown to
have an isotropic velocity dispersion tensor, thus giving a hint on a
dissipational formation process. 
\end{abstract}

\end{opening}

\section{Introduction}

The study of S0 galaxies could provide the link between
the dynamically hot, pressure-supported elliptical galaxies, sustained
by the anisotropy of their velocity dispersion tensor, and the colder,
isotropic and rotationally-supported spiral galaxies.

These dynamically differences are likely to be derived from different
formation histories: under very simple assumptions, we
expect the isotropic structures to derive from the dissipational
collapse of a spherical distribution (or from the mergers of gas-rich
galaxies), and the anisotropic ones to be the result of
dissipationless processes such as mergers of gas-poor parent galaxies.

S0 galaxies shows a very wide range of dynamical behaviors, suggesting
that they might form a somewhat ``mixed'' class at this regard. In
order to investigate the different behaviors of the star kinematics in
these galaxies we developed a self-consistent, multi-component
dynamical model, that could be used to derive simultaneously both the
mass distribution and the mean kinematical anisotropy in S0 galaxies.

Then we looked for a test case, taking data from the literature to
check the model.  We choosed the S0 galaxy NGC~5866, mainly because it
is an object with good photometrical and dynamical data available, and
because velocity and velocity dispersion profiles are present in the
literature also out of the major axis.
These latter profiles are essential to find the anisotropy of the
velocity dispersion tensor of the galaxy.

\section{The hypothesis of the model}

The galaxy is supposed to be composed by the superposition of
different components. For each component, assume:

\begin{itemize}[$\bullet$]
\item There is axial symmetry, so that no bars, warps or misalignments 
      between the different components are allowed;
\item The isodensity surfaces are similar concentric spheroids;
\item The density profile follows a simple $r^{1/4}$ or exponential law;
\item The velocity distribution is locally Gaussian;
\item The anisotropy parameter $\beta$ is constant through the galaxy;
\item The galaxy is rotating around its $z-$axis with velocity $V(R,Z)$,
      consistent with the self-gravity hypothesis;
\item Each component has a constant $M/L$ ratio;
\item Since we are dealing with the inner kinematical properties of the 
      galaxies, we deliberately discard any dark matter contribution to 
      the overall potential, to limit the number of free parameters.
\end{itemize}

We separately fit the major and minor axis profile of the photometry;
the usual ellipse fitting to the 2-dimensional photometric profile is
shown to give worst results here, because in the special case of S0s
the shape of the composite (bulge + disc) isophotes is far from being
ellipsoidal.

Different mass and potential models are then produced with different M/L 
ratios, all compatible with the observed photometric data. 

For each model, we integrate the Jeans equation under the hypothesis
sketched above, obtaining a self-consistent model of the rotational
velocity and velocity dispersion of the different components,
including the asymmetric drift effects.

Assuming that each component has a Gaussian velocity distribution, we
calculate the momenta of the global velocity distribution and
integrate them along the line-of-sight. The stellar line-of-sight
velocity distribution parameters are obtained in the framework of the
usual Gauss-Hermite expansion series \cite{vdf93,gerhard93}

Since we do not actually integrate along the line-of-sight the whole
line shape, but only the momenta of the velocity distribution, the
$V$, $\sigma$, $h_3$ and $h_4$ parameters are obtained by means of a
first order approximation (as described in \inlinecite{vdf93} )

\section{Results and discussion}

\begin{figure}[H]
\psfig{figure=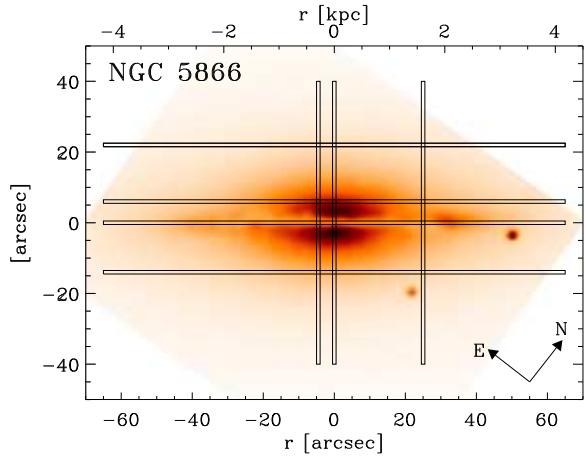,angle=90,width=11.5cm}
\caption{ An image of the galaxy NGC 5866. The approximate position and 
extension of the kinematical profiles \protect\cite{fisher97,KI82}
are shown superimposed on the multi-color image of
the galaxy \protect\cite{pelbal97}. }
\label{fig:slit}
\end{figure}

The model has been checked on the following dataset, available in the literature for the S0 galaxy NGC~5866:

\begin{itemize}[$\bullet$]
\item velocity dispersion and rotational velocity profiles over 7 different 
      axis across the galaxy \cite{KI82,fisher97}, as shown in Fig.~\ref{fig:slit};
\item $h_3$ and $h_4$ profiles for 5 of these axis 
      (not reproduced here for shortness, \inlinecite{fisher97} );
\item photometric profiles along the major and minor axis \cite{pelbal97}.
\end{itemize}

\begin{table}[H]
\caption{Parameters from the Dynamical Model for NGC 5866. 
The labels $b$ and $d$ refers to the bulge and disk components.  }
\label{tab:parameters}
\begin{tabular}{ccccccccc} \hline
\multicolumn{2}{c}{scale radius} 	&
\multicolumn{2}{c}{axial ratio}  	&
\multicolumn{2}{c}{Luminosity} 		&
\multicolumn{2}{c}{Mass [10$^{10} M_\odot$]}&
\multicolumn{1}{c}{} 		 	
					\\ 
\rcline{1-2} \rlcline{3-4}  \rlcline{5-6} \lcline{7-8} 

\multicolumn{1}{c}{r$_b$} 		&
\multicolumn{1}{c}{r$_d$} 		&
\multicolumn{1}{c}{(b/a)$_b$} 		&
\multicolumn{1}{c}{(b/a)$_d$} 		&
\multicolumn{1}{c}{L$_b$} 		&
\multicolumn{1}{c}{L$_d$} 		&
\multicolumn{1}{c}{M$_b$} 		&
\multicolumn{1}{c}{M$_d$} 		&
\multicolumn{1}{c}{i} 
					\\
\hline
$35^{\prime\prime}$ & $ 15^{\prime\prime}$ & 0.8 & 0.15 &
62 \% & 38 \% & 5.58 & 1.72 & $71^\circ$ \\
\hline
\end{tabular} 
\end{table}

\begin{figure}[H]
\begin{center}
\psfig{figure=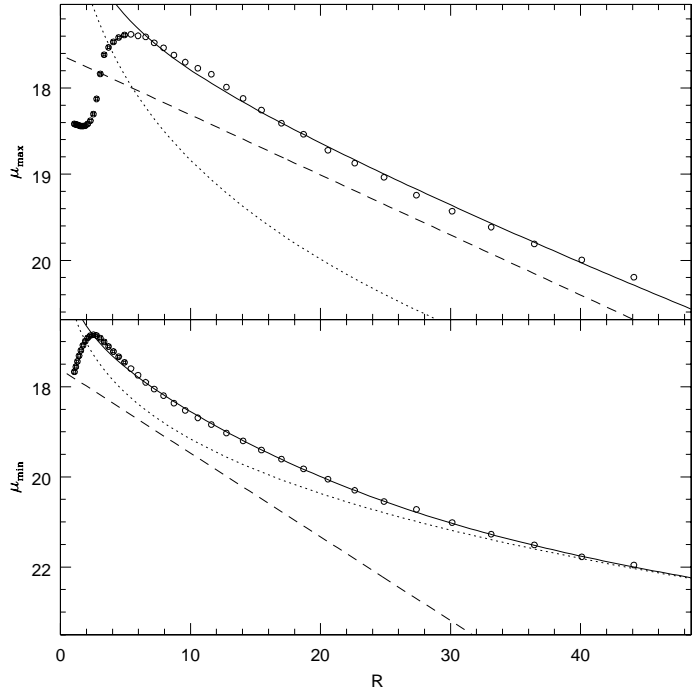,width=10cm}
\caption{ Major and minor axis R-band photometric profiles, compared with 
the two components best-fit model. The dashed lines shown the contribution of 
the bulge and disk to the total surface brightness profiles (solid line). 
The crossed circle represent the points which have been excluded 
from the automatic fit due to the effect of the absorption of the dust lane.}
\label{fig:photo}
\end{center}
\end{figure}

\begin{figure}[H]
\centerline{\psfig{figure=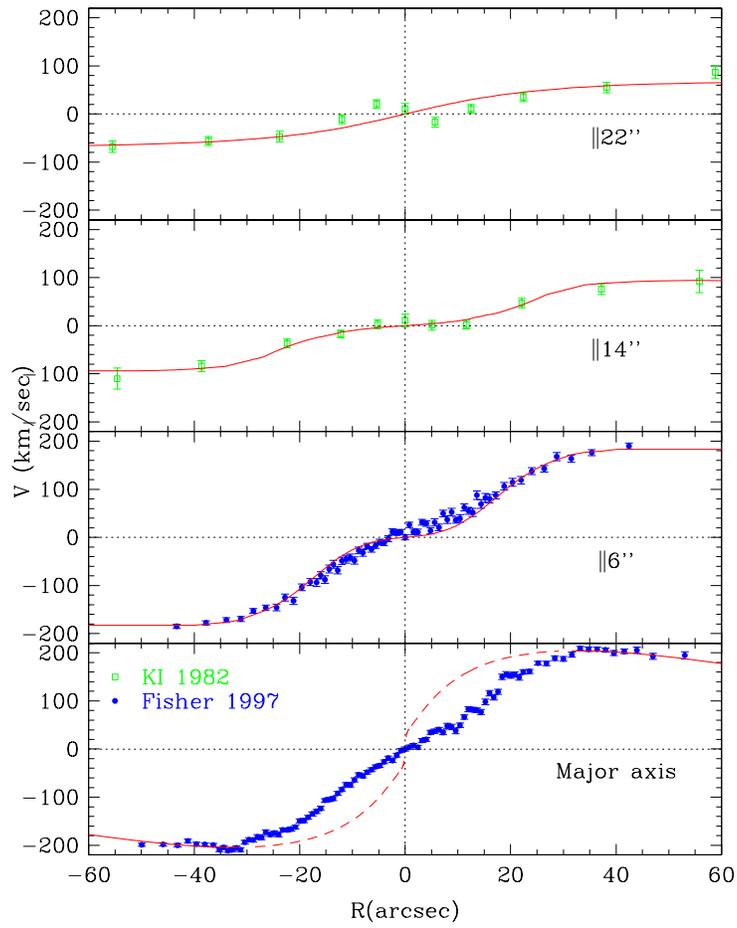,width=10cm}}
\caption{ Rotational velocity profile of stars and gas as compared to the 
model of the stellar kinematics, for the axis parallel to the major axis. The 
dashed line on the model correspond to the region obscured by the dust lane. }
\label{fig:vmaj}
\end{figure}

The presence of  a dust lane, which prevent  us  from looking through
the entire  line-of-sight- of the galaxy,  has been taken into account
by:

\begin{itemize}[$\bullet$]
\item Removing the inner 5 arcsec points from the photometric R-band profile 
(where the sudden drop of the surface brightness profile, in 
Fig.~\ref{fig:photo});
\item Removing from the fit the inner rotational velocities, heavily affected 
      by the dust ( corresponding to the dashed lines models in 
Fig.~\ref{fig:vmaj}).
\end{itemize}
Little effect is expected on the velocity dispersion.

The best-fit model is a two-component (bulge+disk) model whose
parameters are shown in Tab.~\ref{tab:parameters}. The agreement between
the model and the data is good for both the velocity dispersion
profiles and the rotational velocities everywhere, {\em except in the
inner regions of the major axis}. This discrepancy, together with the
somewhat smaller discrepancy in the minor axis rotational profile near
the nucleus, can be easily ascribed to the presence of the dust lane.

\begin{figure}[H]
\centerline{\psfig{figure=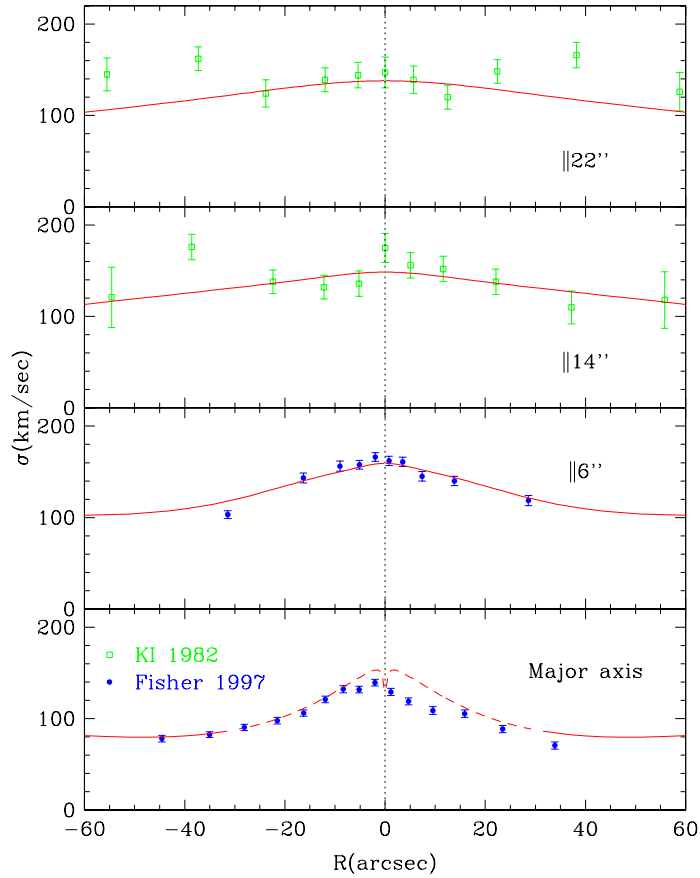,width=9.5cm}}
\caption{ Same as Fig.~\protect\ref{fig:vmaj}, but for the velocity dispersion profiles. }
\label{fig:sigmaj}
\end{figure}

\begin{figure}[H]
\centerline{\psfig{figure=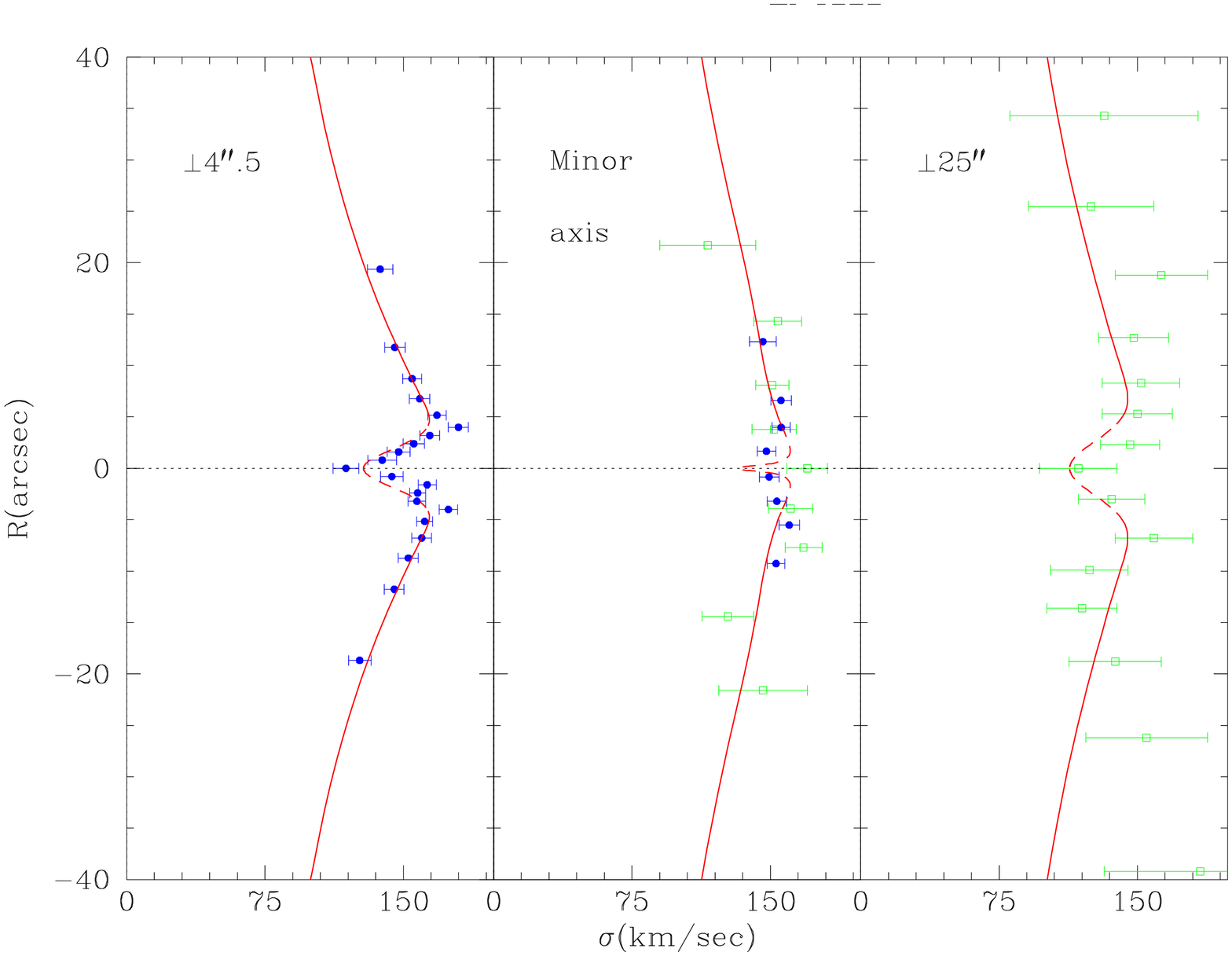,width=8.5cm,angle=270}}
\caption{ Same as in Fig.~\protect\ref{fig:sigmaj}, but for the perpendicular 
cut shown in Fig.~\protect\ref{fig:slit}.  }
\label{fig:sigmin}
\end{figure}

Due to the large uncertainties in the inner structures, we find that
there are several different models that fits the data. However, the
differences are mainly restricted to the inner regions of the galaxy,
while the global parameters of the model are fairly constrained. The
model we are showing in Figg.~\ref{fig:vmaj}--\ref{fig:sigmin} has a
central spherical component that accounts for the big asymmetric drift
noticed in the stellar kinematics, and an isotropic velocity dispersion tensor.

On the other hand, no model with any significant amount of anisotropy
has been able to improve the fit between the model and the data.

\bibliographystyle{klunamed}
\bibliography{evogal}

\begin{figure}[H]
\centerline{\psfig{figure=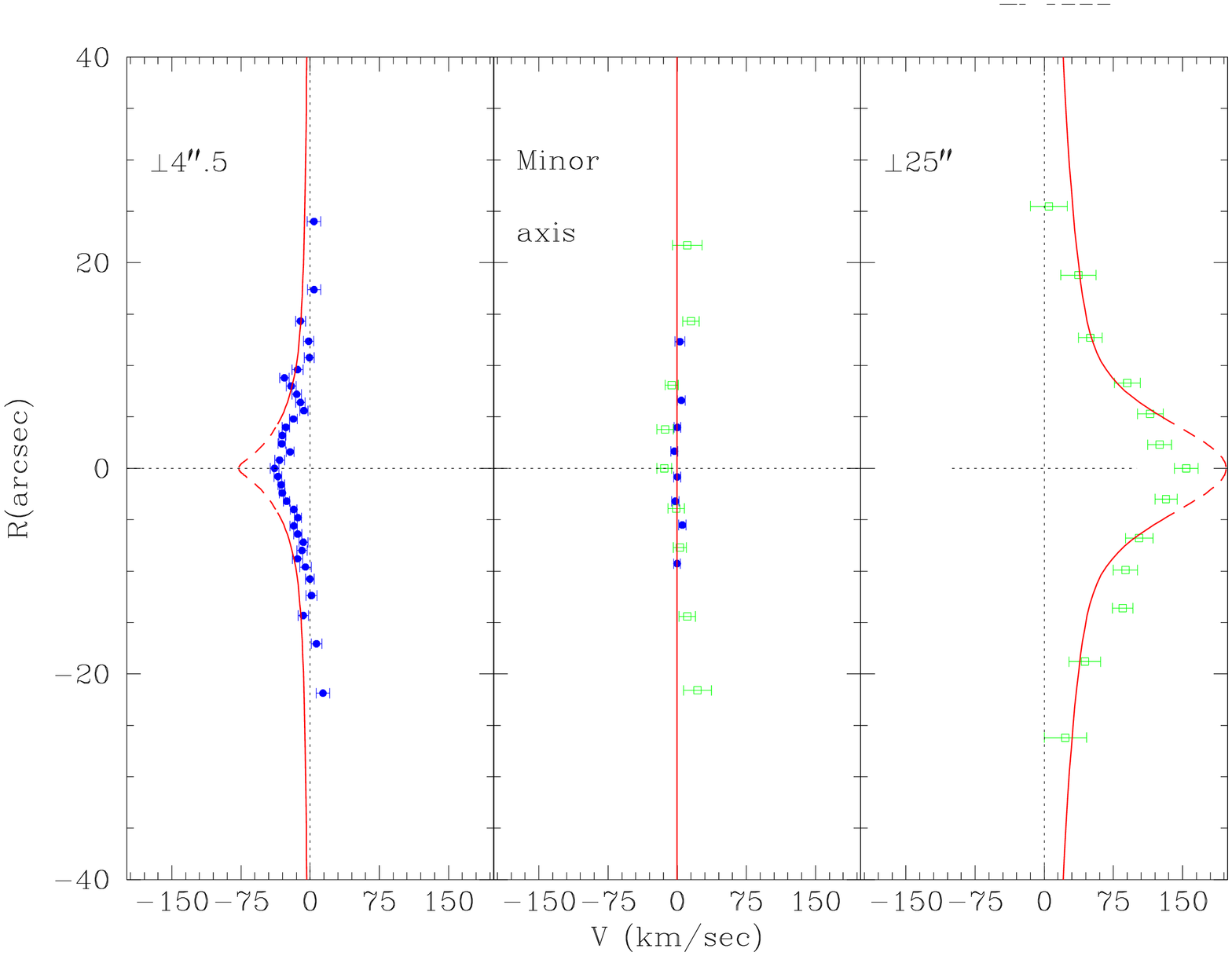,width=8.5cm,angle=270}}
\caption{ Same as in Fig.~\protect\ref{fig:vmaj}, but for the perpendicular 
cut shown in Fig.~\protect\ref{fig:slit}.  }
\label{fig:vmin}
\end{figure}

\end{article}

\end{document}